\PassOptionsToPackage{svgnames}{xcolor}
\documentclass[journal]{vgtc}                     

\onlineid{1425}

\vgtccategory{Research}

\vgtcpapertype{Applications}

\graphicspath{{figs/}{figures/}{pictures/}{images/}{./}}

\usepackage{tabu}                      
\usepackage{booktabs}                  
\usepackage{lipsum}                    
\usepackage{mwe}                       
\usepackage{enumitem}
\usepackage[T1]{fontenc}
\usepackage{mathptmx}                  
\usepackage{svg}
\usepackage{amsfonts}
\usepackage{algorithm}
\usepackage{algorithmicx}
\usepackage{algpseudocode}

\algrenewcommand\algorithmicrequire{\textbf{Input:}}
\algrenewcommand\algorithmicensure{\textbf{Output:}}

\newcommand{\rev}[2]{\textcolor{black}{#2}}
\newcommand{\short}[1]{\textcolor{black}{#1}}
\definecolor{hwcolor}{RGB}{90,180,172}
\definecolor{trueblue}{HTML}{0073cf}

\newcommand{\toolname}{DG Comics\xspace}

\title{\toolname: Semi-Automatically Authoring\\ Graph Comics for Dynamic Graphs}

\author{%
  \authororcid{Joohee Kim}{0000-0002-0745-2339},
  \authororcid{Hyunwook Lee}{0000-0002-5506-7347},
  Duc M.\ Nguyen,
  \authororcid{Minjeong Shin}{0000-0001-6516-0433},
  \authororcid{Bum Chul Kwon}{0000-0002-9391-6274},\\
  \authororcid{Sungahn Ko$^*$}{0000-0002-7410-5652}, and
  \authororcid{Niklas Elmqvist}{0000-0001-5805-5301}
}
\authorfooter{
 \item[]{$^{*}$Corresponding author}
 \item Joohee Kim, Hyunwook Lee, Duc M.\ Nguyen, and Sungahn Ko are with UNIST (Ulsan National Institute of Science and Technology).\\E-mail: \{\href{mailto:joohee@unist.ac.kr}{joohee}, \href{mailto:gusdnr0916@unist.ac.kr}{gusdnr0916}, \href{mailto:ducnm@unist.ac.kr}{ducnm}, \href{mailto:sako@unist.ac.kr}{sako}\}@unist.ac.kr
 \item Bum Chul Kwon is with IBM Research. E-mail: \href{mailto:bumchul.kwon@us.ibm.com}{bumchul.kwon@us.ibm.com}
 \item Niklas Elmqvist is with Aarhus University. E-mail: \href{mailto:elm@cs.au.dk}{elm@cs.au.dk}
}

\shortauthortitle{DG Comics}

\abstract{%
    Comics are an effective method for sequential data-driven storytelling, especially for dynamic graphs\rev{R1-1}{---graphs whose vertices and edges change over time}.
    However, manually creating such comics is currently time-consuming, complex, and error-prone.
    In this paper, we propose \textsc{\toolname}, a novel comic authoring tool for dynamic graphs that allows users to semi-automatically build and annotate comics.
    The tool uses a newly developed hierarchical clustering algorithm to segment consecutive snapshots of dynamic graphs while preserving their chronological order.
    It also presents rich information on both individuals and communities extracted from dynamic graphs in multiple views, where users can explore dynamic graphs and choose what to tell in comics. 
    For evaluation, we provide an example and report the results of a user study and an expert review. 
}


\teaser{
 \centering
 \includegraphics[width=0.95\linewidth]{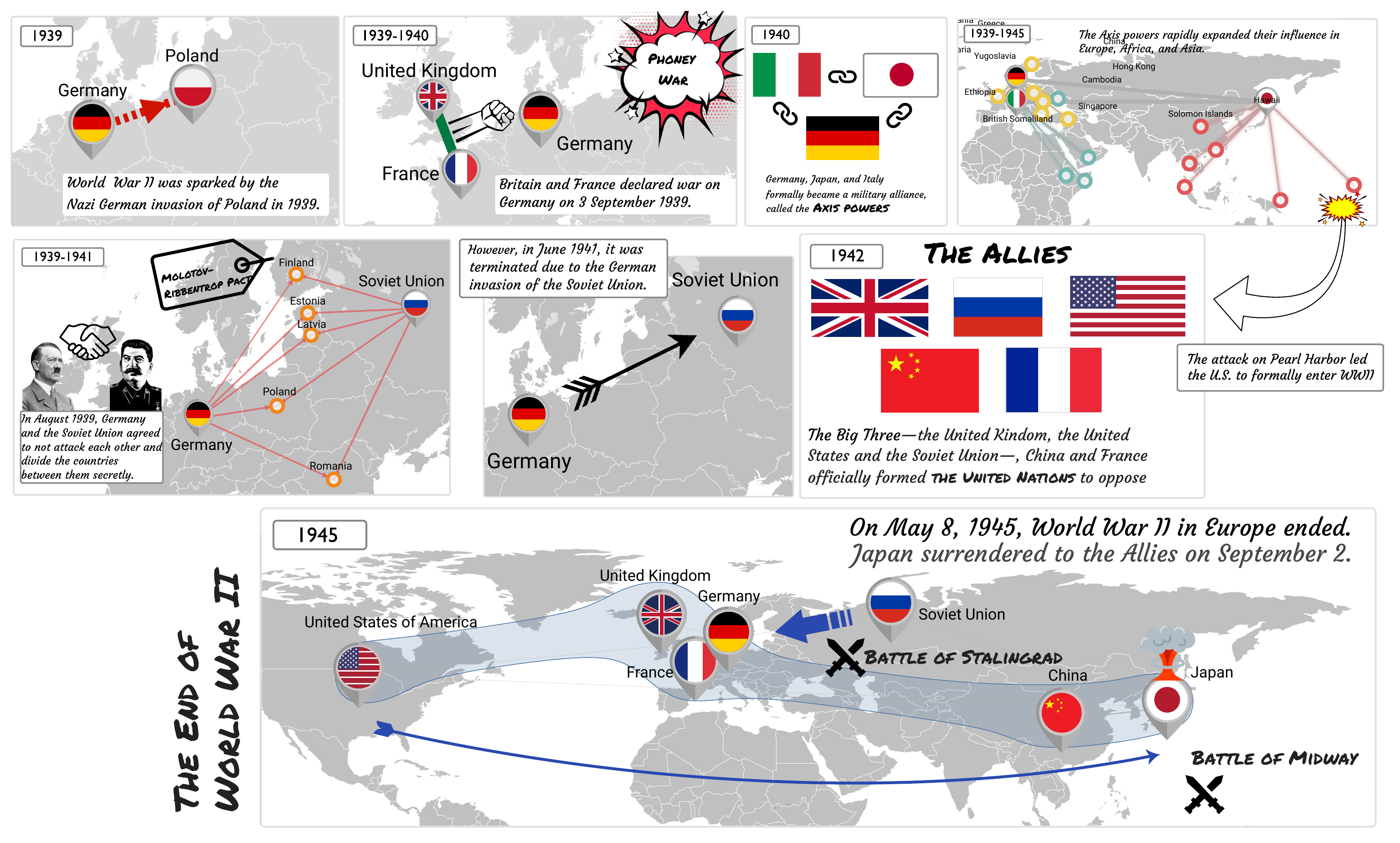}
 \caption{\textbf{World War II.}
 \rev{R1-2}{Graph comic on the history of World War II, covering the causes and significant events, such as the formation of alliances and major battles}, created using \toolname.
 \rev{R1-12}{Note that while the underlying dynamic graph comic was generated automatically, the \toolname tool provides functionality for the designer to manually move nodes, add visual highlighting, and insert the geographic map backgrounds.}
 We used contemporary national flags for easier recognition wherever feasible. 
 }
 \label{fig:teaser}
}

\begin{document}
\maketitle
\section{Introduction}
\label{sec:intro}

Dynamic systems are prevalent in both nature and society.
Catalysts facilitate chemical reactions, while species interact throughout evolution. Scientists collaborate with colleagues and students across various stages of their careers. 
Social relationships form, evolve, and dissolve as individuals make friends, have children, and experience rifts or pass away.
Modeling these phenomena as \textit{dynamic graphs}, 
\rev{R1-3}{where nodes represent entities and links represent their evolving relationships over time}, is a valuable tool for understanding such systems.
However, due to inherent complexity and scale, it is challenging to communicate stories extracted from dynamic graphs succinctly and accurately.
In previous studies, \textit{graph comics}, a comic-based storytelling medium that consists of graph visualizations, is found to be effective in narrating stories involving dynamic graphs~\cite{Zhao2015, Bach16, Segel2010}.
Despite the potential benefits, it is time-consuming for users to create such graph comics based on their data, especially given the lack of dedicated authoring tools.

We address this gap by proposing an approach to automatically generate graph comics from hierarchical clustering that preserves the temporal causality (thus \textit{causality-preserving}) of the events.
To validate the approach, we implement an interactive tool called \textsc{\toolname}:

\begin{itemize}
    \item[T1] \textbf{Storytelling Automation:}
    To combat the complexity of dynamic graphs with many relationships spanning a long period of time, \toolname automates the clustering of temporal events into segments using causality-preserving hierarchical aggregation~\cite{Elmqvist10}.
    \item[T2] \textbf{Storytelling Agency:}
    To facilitate the creative and functional agency of the analyst authoring the story, \toolname yields control of the aggregation level, the main and supporting characters to display, as well as style and formatting choices to the analyst.    
\end{itemize}

\rev{R1-4}{\toolname offers a causality-preserving clustering of graph snapshots in a dendrogram, which users can leverage to initiate graph comic generation.
From a chosen aggregation level, which determines the number of panels, \toolname automatically creates a comic template with a computed layout.
Each panel features the main character(s) undergoing the most significant changes and their relationships.
These changes are depicted with distinct visual representations and captions generated using a template-based approach. 
Users can further develop the template using various editing functions and build original stories through interaction with additional views and tables.
We verified \toolname's usefulness and versatility through (1) an example using a VIS coauthorship network, (2) a user study involving 13 university participants designing graph comics with an international trade network dataset, and (3) interviews with six experts from various domains.} 

\rev{R3-3}{The contributions of this work include (1) robust design requirements derived from challenges identified in prior research; (2) development of a causality-preserving clustering technique with enhanced graph similarity computation as the backbone of automation; (3) implementation of a novel interactive comic authoring tool; and (4) results from an example scenario, a user study, and expert reviews.}

\section{Related Work}

Our work lies at the intersection of graph visualization~\cite{Landesberger11}, data-driven storytelling~\cite{Riche18}, and data comics~\cite{Zhao2015}.
Below we review the literature in all these areas and discuss how our work supersedes prior art.

\subsection{Graph Visualization}


\short{A \textit{graph} or a \textit{network} consists of nodes and links, with nodes representing entities (e.g., people, organizations) and links representing relationships (e.g., friendships, alliances).
Graphs are widely used in domains~\cite{Landesberger11} such as transportation, biology, communication, business, and security. 
As their utility increases, so does their size. Combining networks with data like maps (geospatial graphs), time (dynamic graphs)~\cite{Filipov23}, or text adds complexity.
Conventional graph visualizations include two methods: node-link diagrams~\cite{Ghani11}, which use lines to connect nodes and visual channels (e.g., color, size, line thickness) to distinguish attributes, \rev{R2-4}{and} adjacency matrices~\cite{Behrisch16, Henry06, Henry07b}, which organize nodes in a grid and display connections at intersecting cells.
Prior art~\cite{Landesberger11, McGee19, Nobre19} provides detailed descriptions.}

\short{As graphs grow, visualizations become complex and cluttered. NetworkNarratives~\cite{Li2023NetworkNarratives} provides semi-automated guided data tours to facilitate the navigation of complex networks.
Symbolic representations are one way to overcome this issue by aggregating or hiding nodes. Dunne and Shneiderman~\cite{DBLP:conf/chi/DunneS13} propose a simplification technique that uses fan, connector, and clique motifs to save space and improve understanding of large graphs. To prevent misunderstanding and address the complexity of large graphs, some research~\cite{Henry06, Henry07b} combines node-link diagrams and adjacency matrices, enabling efficient exploration of networks. Yoghourdjian et al.~\cite{Yoghourdjian18} introduce graph thumbnails for high-level structure visualization, allowing easy identification, comparison, and overview of multiple large graphs. 
Ghani et al.~\cite{Ghani11} use dynamic insets to show off-screen node neighborhoods, while May et al.~\cite{May12} improve off-screen awareness by providing graph neighbor information. 
Visualizing groups or clusters in graphs is another area of research. Saket et al.~\cite{Saket14} propose a taxonomy for graph groups, enumerating tasks such as group-only, group-node, group-link, and group-network. Vehlow et al.~\cite{Vehlow17} survey techniques for visually presenting graph groups, categorizing them into visual node attributes, juxtaposed, superimposed, and embedded methods. 
}

\subsection{Visualizing Dynamic Graphs}

\short{Dynamic graphs are graphs whose relations among entities change over time. 
Such changes bring challenges in tasks, visualization, and evaluation, prompting significant research efforts.
Ahn et al.~\cite{Ahn14} categorize temporal features of dynamic graphs by the rate and shape of changes and individual events.
Beck et al.~\cite{Beck17} survey dynamic graph visualization techniques, focusing on presentation methods such as animated diagrams and static timeline-based charts.
Analyses of temporal features from individual node/link, group, and network perspectives inspired how \toolname presents information on dynamic graphs. Numerous techniques have been developed for visualizing dynamic graphs. Elzen et al.~\cite{Elzen14} visualize graph evolution with sequence views, and Burch et al.~\cite{Burch17} propose pixel-oriented visualizations. GeneaQuilts~\cite{DBLP:journals/tvcg/BezerianosDFBW10} use a hybrid adjacency matrix and node-link representation to visualize family trees over time.  Bach et al.~\cite{Bach14, Bach15} employ 3D matrix cubes and small multiples of adjacency matrices.}


\short{To reduce the complexity of dynamic graphs, computational methods like spectral graph wavelets~\cite{Col18} and diachronic node embedding~\cite{Xu20} hierarchically aggregate~\cite{Elmqvist10} graph snapshots based on graph structures~\cite{Archarmbault08} or attributes~\cite{Hadlak13}. The computation results effectively visualize the changes by time (e.g., small multiples~\cite{Bach15}). 
For example, Elzen et al.~\cite{Elzen16} present a novel visual analytics pipeline that allows users to track graph changes with points. 
The pipeline consists of democratization, vectorization and normalization, dimensional reduction, and visualization, transforming graph snapshots into points. 
Cakmak et al.~\cite{Cakmak21} propose multi-scale snapshot visualization with graph2vec~\cite{Narayanan17}, creating temporal summaries of dynamics graphs.
We refer to~\cite{Xia21} for an extensive survey result on graph learning algorithms.}

\subsection{Graph Comics and Authoring Tools}

\short{Comics~\cite{McCloud1994} are a storytelling genre that presents stories with combinations of illustrations, text, and annotations on various layouts.
Segel and Heer propose using comics for data-driven storytelling in their seminal work on narrative visualization~\cite{Segel2010}.
Data comics, an emerging medium of data-driven narrative visualization, leverage the visual language of comics, including layouts, characters, and captions~\cite{Zhao2015, Bach17}.
Research has focused on identifying characteristics of data comics~\cite{Tong18}, developing authoring tools~\cite{Zhao2015, Chen2023}, and specific applications such as visualization education~\cite{Wang19b} and user study reports~\cite{Wang21}.
Bach et al.~\cite{Bach16} conduct a design study on storytelling with dynamic networks, proposing design factors:
visual representation of graph elements and changes, temporality of changes, element identity, cast of characters, level of detail, overview and detail, and representation of multivariate networks.
They also define a design space for data comics by analyzing common patterns in existing storytelling media (e.g., infographics, data videos)~\cite{Bach18}. 
This design space has two dimensions: content relation patterns (e.g., narrative, temporal, faceting, visual encoding, granular, spatial patterns) and panel layout (e.g., linear, tiled, parallel, grid).
Wang et al.~\cite{Wang19} conduct controlled and in-the-wild user studies to investigate the benefits of data comics compared to infographics. 
Their experiments reveal that data comics improve understanding and recall of information, and are preferred for enjoyment, focus, and engagement.}

\short{Several data comics authoring tools have been proposed for various computing environments (e.g., tablets, computational notebooks, coding environments).
DataToon~\cite{Kim19} is the first effort to help users design comics on dynamic graphs with pen and touch interaction. 
Computational notebooks are a new tool for analyzing, visualizing, and sharing datasets.
However, their results, which combine programming code, notes, and analysis, often struggle to communicate with the audience.
To address this, Kang et al.~\cite{Kang21} propose ToonNote, an extension that converts notebooks into data comics.
CodeToon~\cite{Suh22}, similar to ToonNote in motivation, focuses more on coding environments, facilitating code-aligned storytelling and automated comic generation. Most data comics are static to guide readers through a specific flow and layout.
Wang et al.~\cite{Wang22}, posing questions on the interactivity of data comics, formalize operations for interactive comics (e.g., content highlighting, panel addition/removal).
They also present a lightweight scripting approach with six goals: navigation, details on demand, changing perspective, branching, pause and reveal, and input data.}

\subsection{Comparison to Prior Art}

Our proposed work in this paper uses comics to visualize dynamic graphs changing over time, and our approach is novel over the literature:

\begin{itemize}
    \item Building on Segel and Heer~\cite{Segel2010}'s taxonomy, Zhao et al.~\cite{Zhao2015} first proposed data comics for narrative visualization, but our approach goes beyond their work by applying the idea to graphs.
    
    \item While Bach et al.~\cite{Bach16} apply comics to graphs, our work proposes automatic graph comic generation as well as an interactive editor.
    
    \item Authoring tools for data comics exist~\cite{Kim19, Kang21, Suh22, Zhao2015}, but ours is unique to dynamic graphs and semi-automatically builds comics.

    \item Prior work visualizes temporal summaries of dynamic graphs~\cite{Bach15, Cakmak21, Zhao:2015}, but we apply the idea to automating data comics.

\end{itemize}

\section{Challenges and Design Requirements}
\label{sec:requirements}

The aim of this paper is to develop an authoring tool for data comics that automatically generates a narrative in comic form and enables users to efficiently reorganize them into a coherent storyline. We accomplish this by adhering to the two design principles outlined in \cref{sec:intro}: \textit{storytelling automation} versus  \textit{storytelling agency}. These principles guide authors in managing scale and complexity while preserving their creative and expressive vision for the data-driven story.

To achieve these goals, we reviewed existing fundamental design principles formalized for storytelling~\cite{Bach18, Hullman13, Lee15, Segel2010, Kosara13, Stolper16} and data comics~\cite{Zhao2015, Bach16, Bach17, Bach18, Zhao23}.
We also based our work on prior research in dynamic graphs regarding visual analysis~\cite{Kale23, Vehlow17, McGee19, Landesberger11}), tasks~\cite{Filipov23, Ahn14, Beck17}, and computational methods~\cite{Elzen16, Xu20, Col18}, as well as data comics authoring tools~\cite{Kim19, Kang21, Suh22, Wang22}. 
Based on the design principles and literature survey, we derive several challenges, as listed below. 

\paragraph{C1--Size and Complexity.}

Creating comics from dynamic graph data demands identifying key narrative elements, such as pivotal nodes or events, for storytelling.
This task ranges from emphasizing critical nodes to elaborating on connections for specific events.
Viewing dynamic graphs through various lenses, like individual nodes or node communities, can reveal these elements.
However, analyzing dynamic graphs across multiple perspectives and timeframes is challenging due to the size and complexity of general dynamic networks.


\paragraph{C2--Identifying Characters.}

Because data comics are a genre of comics, they follow common design patterns in comics~\cite{McCloud1994}. 
An important pattern is the existence of \textit{main} and \textit{supporting characters}~\cite{Segel2010, Kosara13, Zhao2015, Bach17, Tong18}.
Main characters drive the story while supporting characters play key roles in the narrative.
Identifying which nodes serve as main or supporting characters poses a challenge, given the multitude of nodes with varying and evolving characteristics. 


\paragraph{C3--Changes Over Time.}

The single most important task in a dynamic graph visualization is to show changes \rev{R1-5}{in the graph} over time~\cite{Ahn14}, particularly for the main and supporting characters, \rev{R1-5}{or their relationships.}
The challenge lies in detecting the changes, presenting them, and annotating the story.
Detecting temporal changes requires a deep understanding of the data, and is difficult to do manually, especially for large graphs.
Visualizing such changes is also not a trivial problem.
Finally, annotations are essential for explaining the context or conveying additional insights.


\paragraph{C4--The Language of Comics.}

Comics employ a distinctive narrative structure, visual language, and temporal sequencing to engage readers.
For end-users unfamiliar with these conventions, effectively integrating them into dynamic graph comics can be daunting.
Ensuring these elements are used effectively to convey complex information in an accessible and compelling manner requires a deep understanding of comic artistry and narrative technique.


\paragraph{Design Requirements.}

Below we present design requirements (R1--R5) for our proposed data comics authoring tool that can help users overcome these challenges (C1--C4):

\begin{enumerate}[label=\textbf{R\arabic*}]

    \item Allow users to choose \rev{R1-6}{\textbf{temporal granularities}} for the data comics that reflect their desired levels of detail (C1).
    
    \item Enable \textbf{multi-perspective storytelling} (individual, community-based, metric-based, etc) (C1).
    
    \item Help users \textbf{find main characters and supporters} for initiating a narrative (C2).
    
    \item Support users to \textbf{detect and present temporal changes} of dynamic graphs with appropriate annotations (C3).
    
    \item Provide \textbf{comics authoring mechanisms} to assist users in following genre conventions (C4).
    
\end{enumerate}

\section{Graph Comic Generation \rev{R3-3}{Techniques}}
\label{sec:graph-generation}

\rev{R3-3}{We introduce techniques for streamlining the generation of graph comics.} 
We develop a hierarchical clustering algorithm that groups snapshots into \rev{R1-6}{different temporal granularities} based on their similarity. \rev{R3-3}{This clustering produces a dendrogram, enabling users to automatically create comics by selecting the number of panels.}

\subsection{Visualizing Dynamic Graphs as Comics}
\label{sec:visualization-approach}

A \textit{graph snapshot} is a static graph representing the state of a dynamic graph at a specific point in time.
We can consider a dynamic graph as a collection of graph snapshots: snapshots organized in a temporal sequence, where a snapshot is generated whenever there is a change in the dynamic graph, such as the addition or deletion of a node or link.
Such a dynamic graph can be na\"{i}vely visualized as a data comic by rendering one \textit{comic panel} per graph snapshot and then visualizing the specific snapshot inside its panel as a node-link diagram or adjacency matrix. 
We provide continuity between adjacent panels by freezing the position of nodes from one panel to the next and emphasizing changes---node and link additions and deletions---using visual highlighting, such as colors, callouts, and visual effects.
The result is a \textit{graph comic}~\cite{Bach16}: a sequence of panels showing a graph changing over time.

Unfortunately, this na{\"i}ve approach is not practical for graphs spanning long time periods or involving many changes because the resulting comic will have a prohibitively large number of panels. 
Furthermore, large graphs with many nodes and links will yield panels that are so cluttered that individual changes are hard to spot.
Below we describe practical approaches to address these shortcomings:

\begin{itemize}
    \item\textbf{Long times:} We present a \textit{causality-preserving temporal clustering} algorithm that combines multiple adjacent snapshots into clusters to support balancing the number of panels and the amount of changes in each panel based on the user's needs; and
    \item\textbf{Large graphs:} We propose graph filtering mechanisms based on the concept of \textit{main} and \textit{supporting characters} so that large graphs can be reduced to more manageable subgraphs.
\end{itemize}

\subsection{Causality-Preserving Temporal Clustering}
\label{sec:algorithm}

We present an algorithm that hierarchically groups graph snapshots in a dynamic graph based on the similarity between adjacent snapshots while preserving their causal (temporal) order. 
Instead of showing every graph snapshot, the graph comic can show \textit{snapshot groups} consisting of the union of several temporally adjacent snapshots.
The goal is to facilitate users precisely adapting the number of panels to show the resulting comic, from a single panel representing all changes (i.e., a snapshot group representing \textbf{all} snapshots) to a panel for every individual change in the graph.
We consider four additional requirements for our algorithm:
it must take into account the node and link labels, which are crucial for the graph data;
it should preserve temporal continuity and consistency, meaning it should provide the same similarity for the same input;
it should factor in multiple numerical attributes for both nodes and links; and it should not be computationally heavy.

\subsubsection{Adjacency-based Hierarchical Clustering}

We employ a variant of agglomerative clustering~\cite{Elmqvist10} where, instead of comparing the distance from each graph snapshot to every other snapshot, we only compare the distances between snapshots that are adjacent in time~\cite{Zhao:2015}.
In other words, given a dynamic graph $\mathbb{G}$ consisting of a sequence of graph snapshots $G_t$ for each time $t \in [0 \dots T]$ for the time period $T$, only graph snapshots at $t - 1$ and $t + 1$ are compared (i.e., snapshots just before or after the current snapshot). 
This means that the temporal order is preserved when graph snapshots are aggregated into snapshot groups, thus maintaining causality in the resulting comic.

Conceptually, a \textit{snapshot group} is regarded the same as a snapshot: it has a time span instead of a point in time representing the first and last time stamps for its constituent graph snapshots, and its graph is the union of those constituent graphs.
Importantly, the distance metric is defined similarly for both a snapshot and a snapshot group.

To build a cluster hierarchy of a sequence of graph snapshots (and snapshot groups), we must provide a distance metric $D(G_1, G_2)$ that accepts two adjacent graph snapshots (or groups) $G_1$ and $G_2$.
We agglomerate the dynamic graph sequence by progressively selecting the two adjacent snapshots with the smallest distance and replacing them with a snapshot group combining them. 
This means that the number of snapshots in the sequence will monotonically decrease until a single snapshot group representing all of the original graph snapshots remains.

\begin{figure}[t]
    \centering
    \includegraphics[width=0.9\columnwidth]{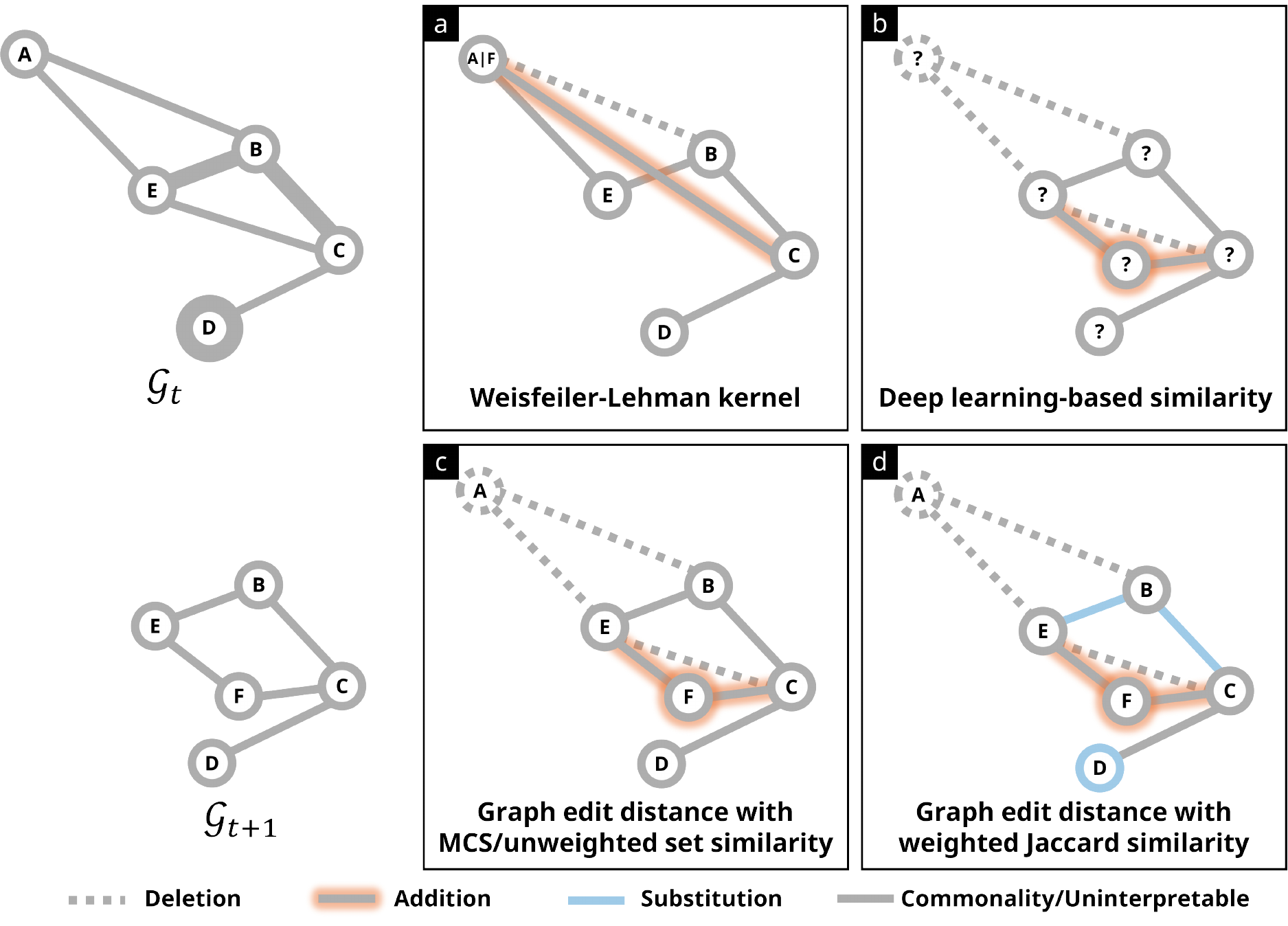}        
    \caption{\textbf{Comparison of graph distance metrics for two graphs.}
    $\mathcal{G}_t$ and $\mathcal{G}_{t+1}$ with different labels and attributes with (a) Weisfeiler-Lehman kernel-based distance, (b) deep learning-based similarity (or distance), (c) graph edit distance with MCS or unweighted set similarity, and (d) graph edit distance with weighted Jaccard similarity (our choice).
    The thickness of the line is proportional to the node or link attributes.}
    \label{fig:algorithm}
    \vspace{-0.7cm}
\end{figure}

\subsubsection{Graph Edit Distance with Weighted Jaccard Similarity}

Several graph similarity measures exist in the literature. 
Basic graph topology measures for unweighted, unlabeled, and undirected graphs are not good metrics of similarity, as real-world dynamic graphs often have both node and link attributes that factor into similarity.
Next, we review existing work on graph distance measure metrics and propose our new distance measure metric used in this work. 

We find four types of metrics: graph edit distance (GED), set similarity, kernel-based measures, and deep learning-based measures. 
GED calculates a weighted summation of predefined graph edit operators, such as insertion, deletion, and substitution of a node or link. 
Although simple and effective, GED requires human effort to determine each edit operator and the corresponding cost. 
Furthermore, existing GED methods define substitution as a link or node replacement, which is not suitable for real-world graphs~\cite{Gao10}. 
For example, as shown in \cref{fig:algorithm} (c), it cannot measure the attribute change of node D. 

Set similarity can be used as a graph similarity metric based on the number of common nodes (or links) in two graphs~\cite{Kogge16, Varma22}.
The main advantage of this method is that it has linear computational complexity and is suitable for real-world applications because of its label awareness.
However, similar to GEDs, it cannot measure attribute changes. 

Kernel-based methods using graph topology are another line of research for graph similarity computation.
Examples include path-based kernels~\cite{Kashima03, Gartner03}, subtree pattern-based graph kernels~\cite{Ramon03, Mahe09}, and Weisfeiler-Lehman optimal assignment kernels~\cite{Kriege16}.
While effective in measuring the overall similarity over the entire graphs, they often omit node attributes (R2) and cannot be used for varying aggregation levels (R2). 
In addition, as shown in \Cref{fig:algorithm} (a), it may lose the label information (e.g., treating link $(A, E)$ and $(E, F)$ as the same one). 

Deep learning-based methods~\cite{Ma21, Yan16} have been developed to overcome the heavy computational costs in measuring graph similarity.
However, the explainability of deep learning methods remains an open problem, so we do not use them in this work, as our graph comics need multi-perspective storytelling (R2) with solid reasoning (R3--R4).

After reviewing existing metrics and their characteristics, we find that traditional methods, including GED and set similarity, are the most appropriate metrics for our purpose. 
However, they cannot be directly used for our work, as they are not able to measure attribute changes.
As such, we develop an enhanced method with GED and weighted Jaccard similarity. 
Our method calculates vector-form Jaccard similarity~\cite{Ruzicka1958} for each common node or link. 
For example, as shown in \cref{fig:algorithm}, we find common elements with labels (e.g., $D$ or $(B,E)$) and then calculate vector-form Jaccard similarity for the attributes of each common element.
As a result, we can obtain the degree of changes per link or node (e.g., change of attributes for node D), as in \cref{fig:algorithm} (d). 

\begin{figure*}[t]
    \centering
    \includegraphics[width=\textwidth]{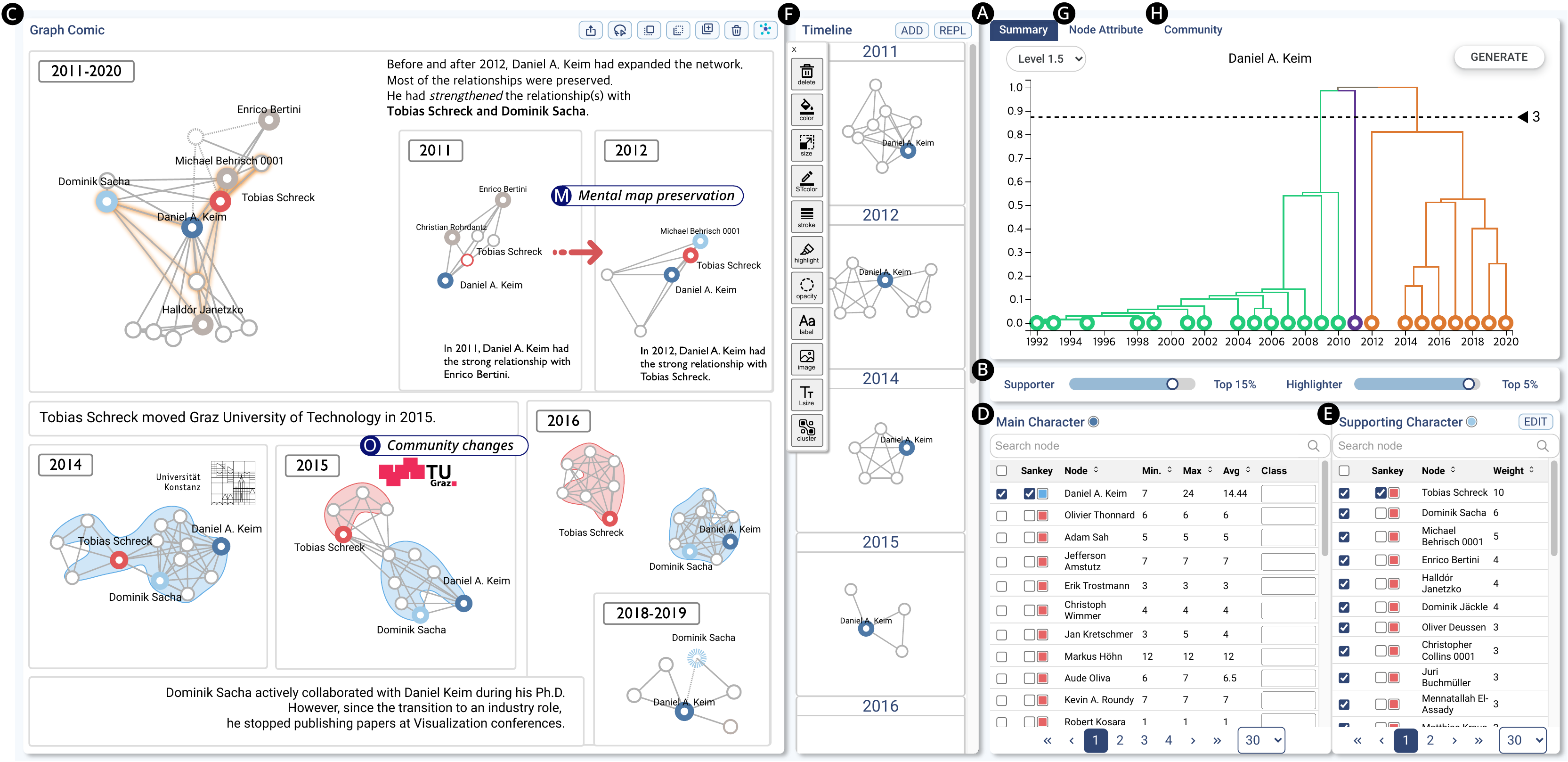}
    \vspace{-0.4cm}
    \caption{\textbf{\toolname overview.}
    \toolname offers 
    (A) a \textit{Summary View,} 
    (B) sliders for filtering and highlighting nodes, 
    (C) a \textit{Graph Comic View}, 
    (D) \textit{Main Character} and 
    (E) \textit{Supporting Character} tables, and 
    (F) a \textit{Timeline View}.
    Users can switch to 
    (G) the \textit{Node Attribute Table} or 
    (H) \textit{Community View} using the tab. 
    It supports (M) \textit{mental map preservation} by fixing nodes across displays, and (O) \textit{community changes} using bubble sets. 
    }
    \label{fig:system}
    \vspace{-0.5cm}
\end{figure*}

\section{The \toolname System}

\short{We design \toolname to meet the challenges and satisfy the requirements from \cref{sec:requirements}.
Our approach outlined in \cref{sec:graph-generation} efficiently manages complexity (R1) for automatic graph comic generation. The algorithm output, visually represented as a dendrogram (\cref{sec:system_overview}), allows users to interactively choose an aggregation level for story fragments, which can then be depicted as comic panels (\cref{sec:comic_view}).
Each comic panel represents a story fragment, and the node-link diagram in the panel visualizes changes during that interval (R4).
Users can build their own stories by inspecting potential main and supporting characters (R3) with graph evaluation metrics (R2), such as node centrality, node degrees, and adjacency (\cref{sec:tables}), as well as individual graphs at different time points (\cref{sec:timeline_view}).
\toolname facilitates the observation of communities (\cref{sec:community}) where characters are involved (R2).}
Finally, it enables editing based on the language of comics (\cref{sec:comic_view}), including fonts, motion lines, captions, and layouts (R5). \rev{R1-10}{\toolname employs Next.js for the frontend, FastAPI for the backend, and the D3 library~\cite{Bostock2011} for visualizations, including computation of node-link diagram layouts.}
\rev{R1-11, R3-4, R3-5}{The source code is available at \href{https://github.com/joohe-e/DGComics.git}{github.com/joohe-e/DGComics.git}.}

\begin{figure}[t]
    \centering
    \includegraphics[width=0.9\columnwidth]{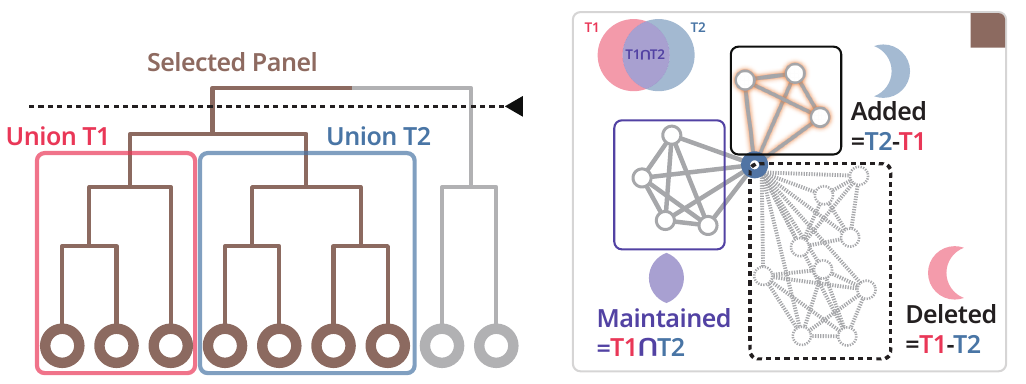}
    \vspace{-0.4cm}
    \caption{\textbf{Data abstraction and presentation.}
    Each cluster below the depth slider (left) represents a panel (right) through a set operation of two subcluster snapshots. The aggregate graph snapshot is computed by the union of graphs for timespans T1 and T2. 
    }    
    \label{fig:set}
\end{figure}
\setlength{\textfloatsep}{5pt}

\subsection{Summary View}
\label{sec:system_overview}

Comics are made of \textit{strips} of one or more \textit{panels}~\cite{McCloud1994, Saraceni2003}, each depicting a portion of the narrative and the sequence typically showing temporal progression.
In a graph comic, each panel contains a snapshot of the dynamic graph sequence (either as an individual or a group) and a caption that provides information about the snapshot.
The \textit{Summary View} (\cref{fig:system}A) facilitates the automatic generation of a series of panels via a dendrogram that visualizes the hierarchical clustering result (\cref{sec:graph-generation}).
Each leaf node of the dendrogram refers to a graph snapshot at a specific time point.
The X-axis indicates the time from the beginning to the end of the data, and the Y-axis notes the normalized similarity between two adjacent snapshots from 0 to 1.
The more similar the groups or individual snapshots are, the lower their position is connected, as the distance is shorter. A horizontal dashed line, called a \textbf{depth slider}, partitions the dendrogram into multiple clusters, visually encoded with different colors. The number beside the line indicates the number of clusters formed at that depth level.

\rev{R1-9}{Users can select the number of panels to include in the comic strip by moving the depth slider (R1). As users adjust the slider, the clusters are colored differently. They can generate the comic strip by clicking the \raisebox{-.2\baselineskip}{\includegraphics[height=1em]{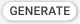}} button. Each branch selected by the depth slider returns the characters that change the most over the timespan corresponding to the cluster (R3).
For example, if a user clicks {\includegraphics[height=1em]{/icons/toolbar/generate}} after positioning the depth slider as shown in \cref{fig:system}A, three panels are created in the comic view, each representing the assigned clusters (green, purple, and orange).}
\toolname offers two options for constructing subgraphs with the ego being the resultant main character(s): a 1.0-level ego network, which includes the ego and its 1-degree alters, and a 1.5-level ego network which also includes the ties between the ego's alters.
Users choose an option based on what relationship they want to present (R2).

\subsection{Graph Comic View}
\label{sec:comic_view}


\rev{R1-10}{The \textit{Graph Comic View} allows users to edit graph comic templates while managing large dynamic graphs efficiently. To reduce visual clutter and enhance readability, we provide an option to keep only the top N percent of alters based on link weights visible. Users can control this level of visibility using \textbf{supporter slider}, and further highlight specific alters with the \textbf{highlighter slider}. Both sliders can be adjusted simultaneously before users begin editing (\cref{fig:system}B).}

\subsubsection{Visual Design}

\short{We create a node-link diagram for each panel as the union of the panel's children (or the snapshot itself for leaf nodes). Each panel visualizes the changes between two consecutive children snapshot groups, highlighting added nodes and links with a neon effect and deletions with dashed strokes (R4).
We abstract the data for each change using set operations: subtracting the former timespan (T1) from the latter (T2) yields the added subgraph, while the reverse reveals the deleted subgraph (\cref{fig:set}).
Supporting characters are defined as nodes with greater weight in their relationship with the main character.}
We emphasize main and supporting characters by thickening the stroke of the nodes, applying diverging colors, and including labels; their colors can be changed using a color picker.


\textbf{Panel Layout.} Since each panel contains a narrative, its size and placement are crucial for conveying the message.
Specifically, viewers can infer event duration from the width of the panel~\cite{Saraceni2003} and determine the chronological order based on the panel's position in the strip, where time proceeds from left to right and top to bottom~\cite{McCloud1994, Saraceni2003}.
\toolname is designed to generate an appropriate panel layout.
\rev{R3-4}{The layout algorithm first calculates the number of rows, or \textit{tiers}, by taking the square root of the total panel number specified by the user, ensuring each panel maintains a minimum height. 
For example, with 16 panels, \toolname assigns 4 tiers. 
It then allocates panels to each tier to balance the sum of their time intervals. 
The comic is partitioned using this structure, with gutters between panels for clarity. 
This approach normalizes time points across tiers, assuming that panels with more time points convey more information, and encodes event duration into panel size~\cite{Bach16}.}


\textbf{Captions.} \toolname provides a caption template for each panel in the main character's point of view to assist storytelling.
Each caption consists of three clauses about a summary, major change, and the most influential relationship.
The summary describes expansions, contractions, and constancy based on the difference between the graphs before and after.
It includes a prepositional phrase in front to indicate the timespan and reference time for a change. 
Then we compare the total number of nodes added, deleted, and preserved, and state one case that is the maximum.
Lastly, we select the node(s) that have the highest link weights and mark whether the main character had obtained, lost, strengthened, or weakened its relationship with those nodes.
A set of clauses is created per main character, but each clause can be combined into one with multiple subjects if different main characters share the same content.
When the panel portrays a particular time point, it simply reports the strongest relationship.

\subsubsection{Interaction Design}
\label{sec:interaction}

\toolname offers diverse editing interactions to enhance flexibility in content creation (R5).
Users can manage the general format of the canvas and graph drawing using the \textbf{toolbar} (\cref{fig:system}C, top-right) in the \textit{Graph Comic View} and adjust the details, such as text and graph styles within the canvas using the \textbf{toolbox} that appears on the right side of \textit{Graph Comic View} when users interact with the element.


\textbf{Canvas Editing.} We define an editable component of the panel as a \textit{canvas}.
\toolname includes three canvas types: graph, text, and image.
An auto-generated panel contains one graph and two text canvases---a caption and temporal information.
Users can move, resize, and rotate a canvas by interacting with the four sides and corners of the canvas, respectively. 
The canvas editing can be done using the toolbar on the upper right corner of the \textit{Graph Comic View} (\cref{fig:system}C).
Clicking the \raisebox{-.2\baselineskip}{\includegraphics[height=1em]{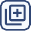}} button reveals a list of canvas types that can be added to the workspace.
One distinct option is \raisebox{-.2\baselineskip}{\includegraphics[height=1em]{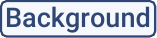}}, which inserts a background image into the graph canvas. To add a graph canvas, users select the timespan on the \textit{Summary View} by brushing before pressing the \raisebox{-.2\baselineskip}{\includegraphics[height=1em]{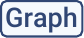}} button.
This will generate the change summary of an extracted main character during the chosen timespan as explained in \cref{sec:visualization-approach}.
The \raisebox{-.2\baselineskip}{\includegraphics[height=1em]{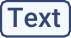}} option enables users to create an empty text box, while the \raisebox{-.2\baselineskip}{\includegraphics[height=1em]{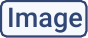}} option offers preset icons such as speech balloons and arrows and supports importing external images via the \raisebox{-.2\baselineskip}{\includegraphics[height=1em]{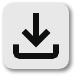}} button.
Users can simply delete, and move back and forth the focused canvas by clicking \raisebox{-.2\baselineskip}{\includegraphics[height=1em]{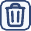}}, \raisebox{-.2\baselineskip}{\includegraphics[height=1em]{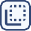}}, and \raisebox{-.2\baselineskip}{\includegraphics[height=1em]{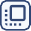}} buttons.


\textbf{Graph Drawing.} \toolname provides multiple modes for generating the graph layout before manually repositioning.
When pressing the \raisebox{-.2\baselineskip}{\includegraphics[height=1em]{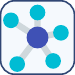}} button on the toolbar, buttons for three modes appear.
The default mode is a \raisebox{-.2\baselineskip}{\includegraphics[height=1em]{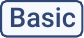}} force-directed graph that uses the concepts of repulsion and attraction to place the relevant nodes closer and irrelevant nodes further.
Derived from this graph structure, the \raisebox{-.2\baselineskip}{\includegraphics[height=1em]{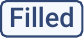}} mode reduces empty space to a minimum in the way of multiplying scale to the relative position.
This mode can be useful to increase the readability of nodes by preventing overlapping. The \raisebox{-.2\baselineskip}{\includegraphics[height=1em]{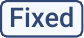}} mode supports mental map preservation by fixing node positions across the panels (\cref{fig:system}M).
To achieve this, we compute a force-directed layout of the union of all nodes and links in the dynamic graph over its full timespan.
This is useful for tracking changes in individual panels and facilitates the easy detection of nodes and links being added or deleted.


\textbf{Text Editing.} When the text canvas gets focused, it triggers a movable toolbox to appear on the upper right side of the \textit{Graph Comic View}.
Users can drag the text content for selection and change it to be bold, italic, or underlined.
The typeface and size are also adjustable.
The background and border of the text canvas can be on and off.
We integrate an LLM (\texttt{Mixtral 8x7B}~\cite{jiang2024}); clicking the \raisebox{-.2\baselineskip}{\includegraphics[width=1.0em]{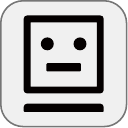}} button will use the LLM to improve the caption automatically.


\begin{figure}[t]
    \centering
    \includegraphics[width=0.85\columnwidth]{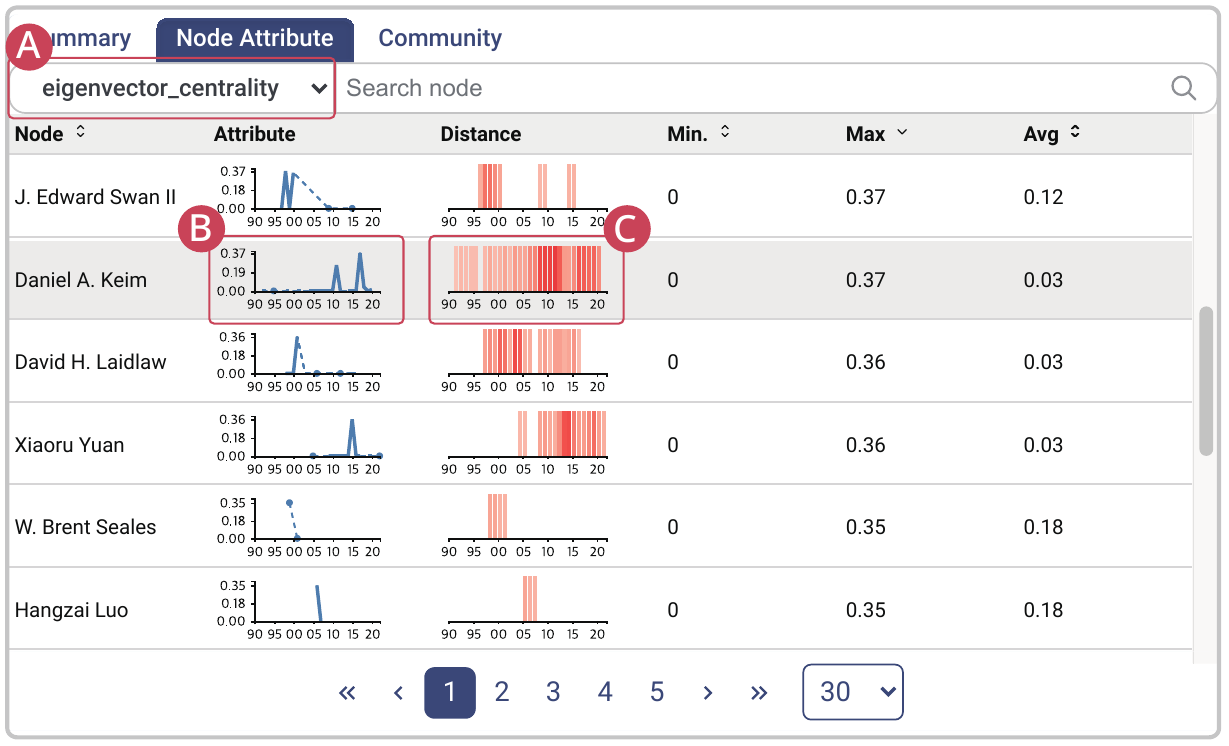}
        \vspace{-0.2cm}
    \caption{\textbf{Node Attribute Table.} 
    This table shows a list of all the nodes and their values over time.}
    \label{fig:filter}
\end{figure}

\textbf{Graph Element Editing.} In the default mode, users can zoom in on the graph canvas by scrolling and pan by dragging the background.
Each node is movable, and its style, along with the link, is editable.
\toolname offers three methods for selecting multiple nodes: users can individually click on the nodes; employ the \raisebox{-.2\baselineskip}{\includegraphics[width=1.0em]{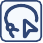}} lasso tool from the graph comic view toolbar for easier group selection; or use a supporting character table by clicking on a character row and then the \raisebox{-.2\baselineskip}{\includegraphics[height=1em]{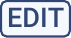}} button (\cref{fig:system}E).
Once the selection is made, group movement of the nodes is available, and a toolbox appears, allowing customization of styles.
The node-level toolbox provides functions for changing color, size, and stroke properties, adjusting opacity, applying highlights, adding labels, inserting images, deleting nodes, and clustering nodes.
Links can be edited by a direct click, and the link-level toolbox offers similar functions, except for labeling, grouping, and image insertion. 
It also features the ability to change markers, particularly useful for directed graphs.
Users can change the color and opacity of a cluster created either by manual grouping or by selecting from the \textit{Community View} (\cref{sec:community}). 
Once all the editing is done, users can export the final product as PNG, JPG, or PDF via the \raisebox{-.2\baselineskip}{\includegraphics[width=1.0em]{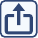}} button.

\subsection{Timeline View}
\label{sec:timeline_view}

We can express temporal changes not only with the symbolic representation but also through the separation of panels.
The \textit{Timeline View} (\cref{fig:system}F) lets users choose a transition style by listing static graphs of different time points.
Users can scrutinize graphs included within the chosen time span and select one to add to the canvas (R4, R5).

The \textit{Timeline View} provides two functions for generating a series of graphs: adding and replacing the selected panel.
The \raisebox{-.2\baselineskip}{\includegraphics[height=1em]{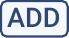}} option enables switching focus from overview to detail by attaching the graphs of certain points in the bottom right corner.
We use the same layout computation for the \raisebox{-.2\baselineskip}{\includegraphics[height=1em]{/icons/toolbar/add}} option to embed multiple panels in a limited space while not covering the root panel completely.
The \raisebox{-.2\baselineskip}{\includegraphics[height=1em]{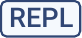}} option replaces the selected panel and arranges new panels in temporal order.

\subsection{Tables}
\label{sec:tables}

We provide three tables in \toolname to select node attributes and characters to display in the graph comic.
The common features of the tables are sorting by values and searching by names.

The \textit{Node Attribute Table} (\cref{fig:filter}) facilitates node exploration and assists in manual main character selection by providing comprehensive attribute values and chart thumbnails (R3).
Users can switch to this table by clicking the second tab above the \textit{Summary View} (\cref{fig:system}G).
The line chart thumbnail tracks the value over time; lines are dashed if data is missing (\cref{fig:filter}B).
To prevent possible bias from inconsistent scales, we also mark minimum, maximum, and average values.
Users can examine different attributes of a node---e.g., the total number of links, PageRank, and centrality---by selecting the option in the top left corner of the view (\cref{fig:filter}A).
The heatmap presents the degree of dissimilarity between the consecutive time points, meaning that the color gets darker as the distance computed with our modified Jaccard index increases (\cref{fig:filter}C).
Using this thumbnail, users can capture moments with significant change.
When selecting a main character on the attribute table, the dendrogram is re-created to reflect the subgraph-level change (R2).
This means that each time node of the dendrogram represents the subgraph grounded in the selected node (\cref{fig:system}A). 

The character tables (\cref{fig:system}D, E) manage the nodes depicted in the graph comic.
The list of nodes is updated whenever users click the graph canvas to focus.
Users can either add or delete nodes by selecting or deselecting the leftmost checkbox.
While the \textit{Main Character Table} (\cref{fig:system}D) shows all the nodes as candidates for the main character, the \textit{Supporting Character Table} (\cref{fig:system}E) lists the neighbor nodes that are linked with the currently selected main character (R3).
The minimum, maximum, and average values in the \textit{Main Character Table} reflect the attribute selected in the \textit{Node Attribute Table}.
The total weight of links to a node in the selected time span is an indication of the importance of the node (\cref{fig:system}E).
The \textit{Class} column in the \textit{Main Character Table} enables direct manipulation of graph element styles.
By entering the tag name as input and clicking on the \textit{Class} header, users can open the CSS editing view to redefine the properties of the corresponding nodes (\cref{fig:system}D).
The \raisebox{-.2\baselineskip}{\includegraphics[height=1em]{/icons/toolbar/edit}} button in the \textit{Supporting Character Table} is explained in Section~\ref{sec:interaction}.


\subsection{Community View}
\label{sec:community}

\begin{figure}[t]
    \centering
    \includegraphics[width=0.85\columnwidth]{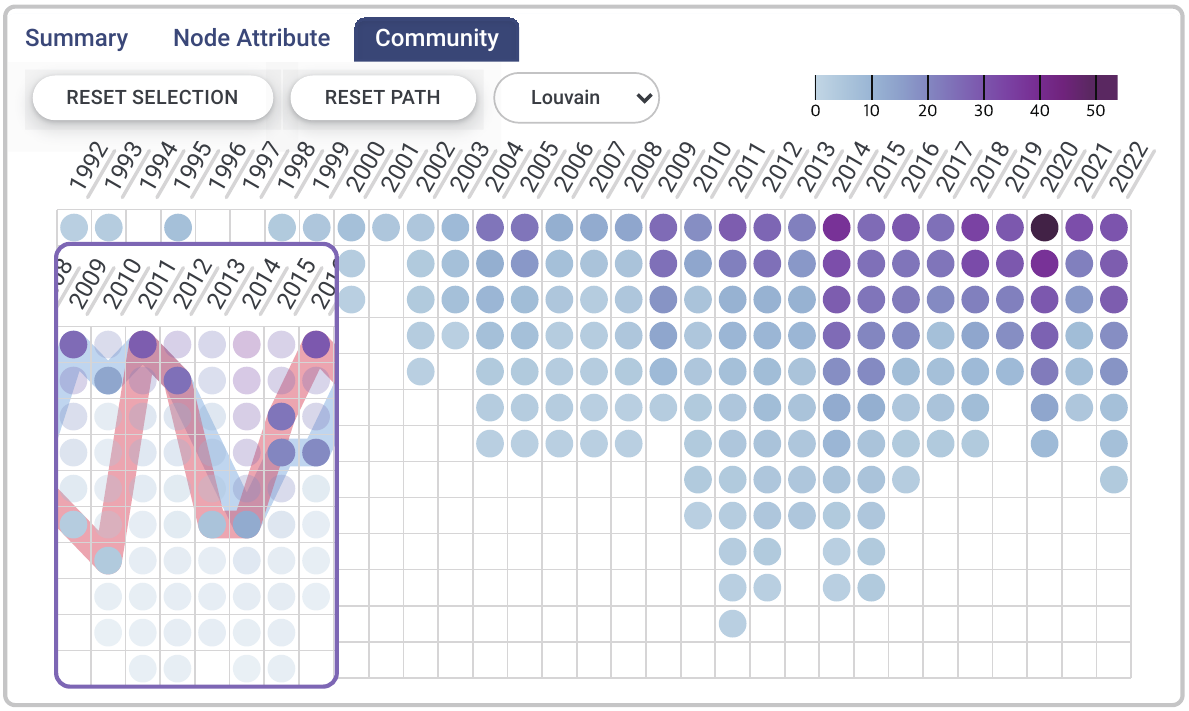}
    \vspace{-0.2cm}
    \caption{\textbf{Community View} changes when communities are selected, as shown in the left inset, where red and blue colors mean Keim and Schreck, respectively, whose communities diverged since 2014.}
    \label{fig:sankey}
\end{figure}

Identifying the communities in which nodes are involved (i.e., community membership) and tracking their evolution is crucial for a deeper understanding of their relationships (R2).
For instance, if two nodes from different communities merge into the same community, we can assume that their relationships become stronger; if they split, the relationship becomes weaker.
A Sankey diagram is a suitable approach to illustrate such flow by representing communities as nodes and transitions between them as links.
To differentiate their nodes and links from graph elements, we call them \textit{community nodes} and \textit{paths}. 
However, its readability significantly decreases with the numerous overlapping paths and varied sizes of community nodes.
Given that the narrative of the graph comic is character-driven, we decided to highlight only the communities of the chosen characters in this work.
Moreover, we redesigned the Sankey diagram to enable fair inspection and comparison of community changes associated with the characters.
\cref{fig:sankey} shows our \textit{Community View}, re-designed in this work. 

In a traditional Sankey diagram, the size of a community node is proportional to the size of the community, making it difficult to observe transitions for characters in smaller communities.
To overcome this obstacle, we encode the community size into a sequential color scheme while normalizing the community node size.
Users can choose a cluster color in a color palette by selecting the second checkbox on the `Sankey' column (\cref{fig:system} D, E).
We set the time as the x-axis and place community nodes from top to bottom based on their community size.
To connect the community nodes for a selected character, we adopt a circular shape with padding inside the grid cell so that the path can enclose it smoothly.
As a result, the \textit{Community View} displays the colored path of community evolution for the selected character while blurring nodes outside of this path as \cref{fig:sankey} shows.

\begin{figure}[t]
    \centering
    \includegraphics[width=0.65\columnwidth]{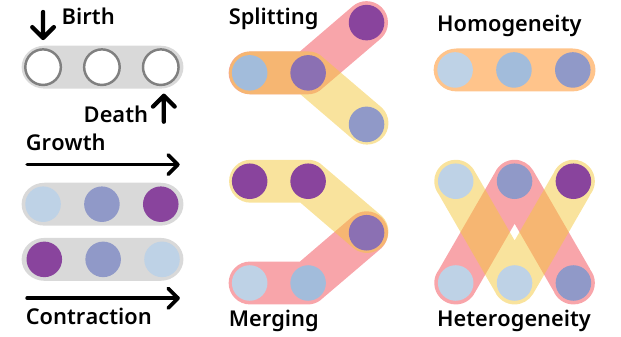}
    \vspace{-0.3cm}
    \caption{\textbf{Community evolution.} 
    Six archetypes of community evolution.}
    \label{fig:com}
\end{figure}

The \textit{Community View} enables users to detect six types of events in community evolution (\cref{fig:com}): growth, contraction, merging, splitting, birth, and death~\cite{Palla07}.
Since \toolname supports character-based exploration for generating narratives, the community is primarily identified by the character.
Therefore, the start and end of the path indicate the birth and death of the character's community within the dataset's time span.
Users can determine growth and contraction through color changes and the vertical position of the circle across different years. 
Furthermore, by selecting multiple characters, users can observe how distinct communities evolve to merge and split at specific time points throughout the time span.
Notably, the visual patterns depicted by multiple paths can illustrate how the communities evolved through various events (e.g., splitting, merging, homogeneity, heterogeneity) over different time periods as \cref{fig:com} shows.
After exploring community evolution, users can add a cluster to the canvas by clicking the community node (\cref{fig:system}O).
\toolname uses Bubble Sets~\cite{Collins09} to highlight the group of characters involved.
The \raisebox{-.2\baselineskip}{\includegraphics[height=1em]{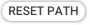}} button (upper left) clears all highlights from the \textit{Community View}, while the \raisebox{-.2\baselineskip}{\includegraphics[height=1em]{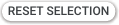}} button removes all the clusters from the graph comic.
The default setting displays the communities detected by the Louvain algorithm, but another option is available in the drop-down list, such as affiliations of authors in the case of Vispubdata.

\section{Example}
We demonstrate the use of \toolname in scientometric analysis with data from the visualization community.
We use Vispubdata~\cite{Isenberg:2017:VMC}, a dataset containing information on 3,620 papers from IEEE Visualization conferences (InfoVis, SciVis, VAST, and VIS) from 1990 to 2022, including conference, year, title, paper type, authors, affiliation, keywords, and number of citations.
\rev{R1-11}{This dataset is frequently used for cross-checking in dynamic network research and is highly relevant to the visualization community. 
To create graph comics with compelling storylines, we constructed author networks, where nodes represent authors and links represent collaborations based on co-authored papers.} 

\subsection{Analysis}
We start with the default dendrogram, which computes the dissimilarity-based hierarchy of the entire dynamic graph, adjusting the depth slider to segment the time into nine divisions. 
This process generates nine panels, offering a visual narrative of the data, which in this case, is the collaboration network in VIS papers. We select 1.5-level ego networks to observe connections between the neighbor nodes as well.
The auto-generated comic reveals \textsc{Daniel A.\ Keim} as a central figure in two distinct periods, 2006--2011 and 2012--2016, with the network expanding in the earlier period and contracting in the latter.
We can now generate a graph comic from Keim's perspective for closer examination.

By opening the \textit{Node Attribute Table} and searching for ``Keim,'' we detect the high dissimilarity (\cref{fig:filter}C) and peak in eigenvector centrality (\cref{fig:filter}B) between 2011 and 2012, which suggests transformations in Keim's collaborative patterns.
With this insight, we create a new dendrogram for Keim and regenerate the graph comic, which unveils two main clusters before and after 2011, indicating major events in the network's evolution (\cref{fig:system}A).
We move the sliders to filter out impactful relationships involving the top 15 percent of total collaborations and highlight the top 5 percent (\cref{fig:system}B).
To examine changes closely, we add the graph canvas by brushing 2011 to 2020 from the \textit{Summary View} and insert the individual graphs of 2011 and 2012 from the \textit{Timeline View} (\cref{fig:system}F).
The tool's features allow us to preserve the mental map (\cref{fig:system}M), helping us to capture the establishment of new relationships and community shifts.
Notably, we observe a strengthening of relationship with \textsc{Tobias Schreck} from 2011 to 2020, prompting us to further compare the dynamics between Keim and Schreck.


In the \textit{Community View}, we color-code Keim's (blue) and Schreck's (red) communities and draw paths to visualize their changes over time (\cref{fig:sankey}).
Despite multiple collaborations, their communities have diverged since 2014 (\cref{fig:system}O), reflecting the impact of Schreck moving to Graz University of Technology in 2015.
This divergence, despite ongoing collaboration, illustrates nuanced dynamics common in academic networks.
Further exploration reveals \textsc{Dominik Sacha} as a significant yet eventually diverging connection, mirroring the evolution of community affiliations and the impact of career milestones such as Ph.D.\ completion and industry employment.
By juxtaposing Sacha's trajectory with Schreck's, we enrich the narrative, offering comparative insights into how individual careers and collaborations shape the broader academic community.
This scenario demonstrates the tool's capacity to dissect, visualize, and narrate complex relational data, making it an invaluable resource for understanding dynamic networks.

\section{Evaluation}
\rev{R1-16}{To explore the usefulness and effectiveness of \toolname, we conducted a comprehensive user study. This study includes usability tests and interviews to gather valuable insights from participants and in-depth feedback from experts to validate its versatile application.}

\subsection{User Study}
We conducted a controlled user study to evaluate \toolname. 
The study consisted of a pre-study questionnaire, a tutorial session, graph comics creation, and a post-questionnaire with interviews. 
We present both quantitative ratings and qualitative insights from the interviews.
We used the COW trade dataset~\cite{Barbieri09, Barbieri16COW} of international trades from 1870 to 2014, covering 207 unique nations.
Using this data, we built a directed graph of international trade where each node and link represents nationality and corresponding trade (in U.S.\ dollars), respectively.

\subsubsection{Participants}

We recruited 13 participants from a university (4 females) aged from 23 to 29 ($M = 25.4, SD = 2.2$).
\rev{R1-14}{The number of participants is consistent with prior work~\cite{Nielsen94, Hwang10, Caine16}, aligning with recommendations for usability testing.}
They were graduate ($n=8$) and undergraduate ($n=5$) students from electrical engineering ($n=4$), computer science ($n=5$), and artificial intelligence ($n=4$). 
All participants had experience in analyzing data using Python and visualizing it with Python, Excel, and PowerPoint. 
Most of them had experience dealing with graph data (e.g., traffic, sensor network, etc.), except for three participants who were aware of graph data but lacked experience in processing it. 
Participants were compensated \$22 (USD) for 2 hours of study. 
The study was reviewed and deemed exempt by our Institutional Review Board.

\subsubsection{Procedure}

The study began with a pre-study questionnaire that included:
(1) demographic information, such as age, gender, and education level;
(2) participant experience with data analysis, data visualization, and graph data; and
(3) participant familiarity with graph comics, the trade dataset, and events of world history relevant to the study. 
We then explained what a data comic is and how it can be applied to graph data. 
We conducted a 20-minute tutorial on how to use \toolname and interpret the visual representations. 
Then participants were given 5 to 10 minutes to explore the system on their own and ask questions about its functions. 
Note that we used Vispubdata~\cite{Isenberg:2017:VMC} for the tutorial session to prevent learning effects on the dataset used in the experiment.
 
After participants felt comfortable with the tool, we introduced them to an international trade dataset spanning from 1930 to 1960.
They were tasked with using \toolname to craft an engaging graph comic, highlighting the evolving relationships between nations.
To aid in their task, we provided access to a Wikipedia page titled \textit{Timelines of Modern History}, which includes sections such as \textit{Timeline of the 20th Century} and \textit{List of Wars}.
The purpose was to accommodate varying levels of familiarity with world history among participants.
While participants could use search engines to gather information and craft their storylines, we prohibited the use of any comprehensive historical resources, such as videos, that might present a finished interpretation of events.

Assistance was offered solely at the request of participants, ensuring that their interaction with the system remained independent.
Participants used Google Chrome in full-screen on two 32-inch monitors with a resolution of $2560\times1600$ pixels---one for information search and the other for the actual comic creation. 
Participants were allotted 1 hour and 10 minutes for this task, during which their interactions with the system were screen-recorded for further analysis.

Following the comic creation phase, we asked participants to fill out a post-study questionnaire.
This questionnaire gauged their opinions on the usability, overall experience, and the helpfulness of different views within \toolname, as well as their satisfaction with their final product, using a 5-point Likert scale.
We concluded each session with a semi-structured interview to explore the reasoning behind questionnaire responses, the narratives constructed, and comprehensive feedback on the system's functionality and user experience.

\begin{figure}[t]
    \centering
    \includegraphics[width=\columnwidth]{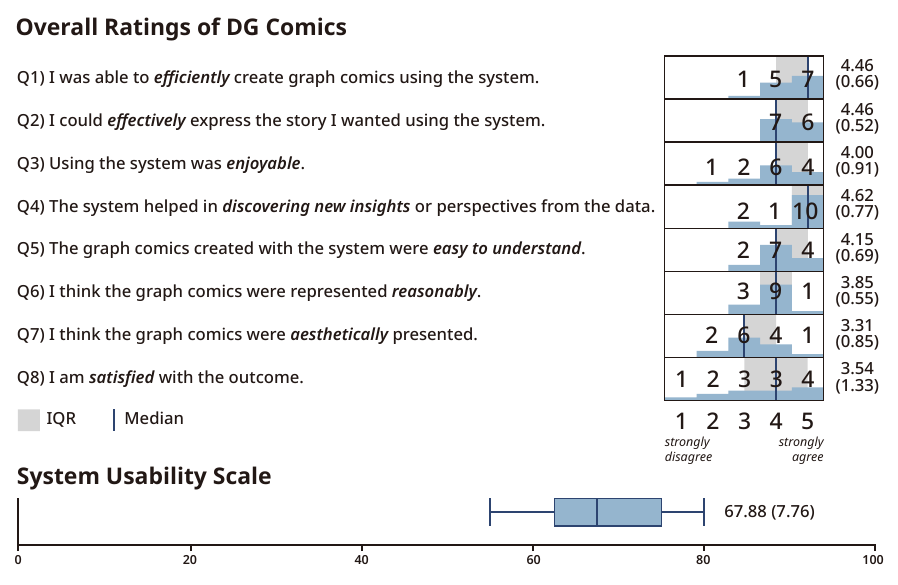}
    \vspace{-0.5cm}
    \caption{\textbf{\toolname ratings.}
    Overall ratings (top) and SUS score (bottom).
    Values are means and standard deviations (in parentheses).}
    \label{fig:rating}
\end{figure}
\setlength{\textfloatsep}{5pt}

\subsubsection{Quantitative Results}

We adopted a 5-point Likert scale to assess the overall experience of the system (Q1-Q4) and the output that participants generated (Q5-Q8).
We used the System Usability Scale (SUS) designed by Brook et al.~\cite{Brooke1996} for global evaluation of systems usability. 
Interpreting the Likert scales as ordinal, we report the quantitative results in a discrete visualization style~\cite{South22}, the histogram~\cite{Zhao23, Huang2022} with medians (as blue lines) and quartiles (in gray areas).
We also provide means and standard deviations considering intervalist views~\cite{Carifio08}.

As \cref{fig:rating} shows, participants felt positive about their experience using the system (Q1-Q3). 
In particular, they highly valued discovering new insights and perspectives from the system (Q4). 
They all agreed that the comics they made were easy to understand (Q5-Q6). 
However, aesthetics and satisfaction with the output varied among the participants, likely due to the discrepancy between their editing skills and expectations toward their outcomes (Q7-Q8).
The SUS score is acceptable ($M=67.88, SD=7.76$); compare this to the average of 68.2 from 273 prior studies~\cite{Bangor2009}.

\subsubsection{Qualitative Results}

We analyzed participants' responses and video recordings associated with the scores. 
Here we present these qualitative results.

\textbf{The summary view supports efficient story generation.}
Most participants reported that the overview with the dendrogram helped generate stories ($MD=4, M=4.15, SD=0.99$). 
Even though most of them had little knowledge of the data, they were able to understand the summary of changes in relationships at a glance.
They used the auto-generated comic as a starting point. 
The hierarchical clustering effectively grouped similar time points to depict relevant stories into one panel, allowing participants to ``\textit{choose how detailed or brief to express the story by adjusting the level} (P5).'' 
Clusters gave a hint to users where to focus; for example, P9 after detecting groups before and after World War II, ``\textit{focused on the specific branches for storytelling the events}'' while P12 ``\textit{scrutinized the clusters as a chunk of changes}.''

\textbf{The community view effectively aids in inferring relationships between nodes and events.}
Participants used the community view to check the relationship between nodes ($n=7$). 
Some participants displayed a node's community to see other nodes in the same cluster and used this information to structure the graph (P3, P7, P8, P11, P12). 
P12 excluded some countries such as the U.S.\ from the graph since ``\textit{their consistency in the relationship with the United Kingdom suggests no special event happening}.’’ 
They instead captured insights by inspecting the community visualization. 
For example, the comparison between the community evolution paths of different countries provides an implication for a significant event when there is a split or a merge. 
P6 searched the interaction between Poland and Germany since ``\textit{their paths overlapped during the 1930s}.’’ 

\textbf{Participants wanted to use the system with their own data.} 
All participants indicated that the system helped them effectively express the story ($MD=4, M=4.46, SD=0.52$). 
Some commented on challenges in visualizing dynamic graphs, especially for big and complex data (P1, P5). 
However, using DG Comics, they were able to convey changes in the relationship between nodes over time regardless of familiarity with the data ($n=7$). 
P7 noted that it became much easier to narrate the story of network evolution because ``\textit{the system helps with the hardest part, [which is] selecting the main character.}''
P10 said that although he did ``\textit{not have an aesthetic sense, [he liked that he could] make a satisfactory product within an hour}.’’ 
Some participants brought up the point that the convenience of the system does not only rely on visualization but also on interpretation of the data. 
``\textit{If the series of time points are clustered, users can think about the reason and assume the circumstance} (P5).’’ 
P1, who conducts research on tracking individuals using ultra-wideband data, said, ``\textit{it would be very helpful to [be able to] detect when interference between two groups occurs}.’’


\subsection{Expert Feedback}
\rev{R1-18}{We conducted online interviews with six experts across various domains, who were invited via emails with consideration of domains and expertise.}
Each expert has a Ph.D. degree and 8 to 19 years of work experience in their respective fields.
Their areas of expertise include Communication (E1: 19 years, E2: 8 years), History (E3: 12 years), Design (E4: 12 years), Management Information System (E5: 8 years), and Economics and Finance (E6: 11 years).
Before interviews, we sent them the information on the interviews (e.g., goal, duration, expected outcomes), a demo video, and a website link to the deployed system.
The interviews began with a short system demonstration, followed by a Q\&A session for any clarifications.
Once interviewees were familiar with the system, we proceeded to discuss its various aspects.


All appreciated the usefulness of the tool in telling a story using an approachable format (i.e., comic).
They reported various strengths of \toolname, including automated story extraction, rich functionality in editing for story presentation, and novelty of the approach in \toolname.
E1 lauded ``\textit{the system is capable of extracting outstanding stories with the corresponding timespan, streamlining the research process for data journalists.}’’
E4 and E5 also noted that laypeople, such as marketers or end-user creators, can use the tool to create and present a story with data that would otherwise have been inaccessible to them.
E1 and E2 raised a similar point, that \toolname enriches users' insights by providing the change in dynamics between characters.
E3 even expressed his willingness to incorporate the system into his curriculum of political history.
``\textit{Given the possible impact of politics on economics, I believe students can gain insights into economic relations, such as international trade, by comparing them with political knowledge through a comic."}
E5 and E6 underlined the contribution of autogenerated output to efficient decision-making. 
Specifically, E5 commented that autogenerated output from dynamic graphs will help users in management who need to analyze dynamically changing relationships among companies---``\textit{an intuitive understanding of supply flows and the impacts of subcontractor changes can help users make strategic decisions in supply chain management.''}
E4 noted that many people need to create and present stories from data but are often frustrated by challenges brought by new and complex data. 
In this situation, he expected the tool would be very helpful. He highlighted the novelty of the approach that \toolname provides, stating,
``\textit{... I give 10 out of 10 for novelty because I have not seen this type of system and support from a complex data storytelling perspective.}''

All experts estimated that the target user spectrum is broad, encompassing a diverse array of individuals. 
With the automatic generation of comic templates and robust editing functions for crafting user-driven stories, experts believe \toolname meets the needs of a wide range of users, from novices to professionals. 
They also commented that usage scenarios with \toolname may vary depending on the user's level of expertise and goals, even among the same expert group. 
For instance, E1 predicted that ``\textit{novice users would primarily utilize the Graph Comic View, whereas experts would navigate between views to gain a deeper understanding of the data.}’’ 
Additionally, E4 mentioned that ``\textit{researchers and data analysts would review the summary of changes from a generated comic to select data before beginning their analysis, yet if they already know the key points, their focus would shift toward authoring to convey their message.}’’

Some experts expressed concerns about the steep learning curve due to the comprehensive features implemented in the system. 
For example, E2 and E4 expressed a concern that this system seems to require some learning time before users fully enjoy the system functions. 
They suggested creating a lite version of the system with fewer features for beginners.
On the other hand, some other experts, namely E2, E5, and E6, indicated that experts may need additional features that help users automatically process unstructured, complex graph datasets.
E5, in particular, mentioned that there is a significant amount of dynamic graph data in companies, but only data scientists can pre-process the data for analysis and presentation purposes---``\textit{If the tool included an interface for processing raw dynamic graph data, it could have a more substantial impact, given the number of employees in enterprises who need to analyze, monitor, and present stories with large graph datasets.}''

\section{Limitations and Future Work}


\toolname provides a flexible and intuitive authoring environment for graph comics creation. 
However, the numerous functions and steps in the authoring process may cause newcomers to face a steep learning curve. 
To mitigate this, providing an interactive tutorial and a help menu that supports keyword search and mouse-over explanations can be a reasonable solution. 
We also consider providing a function for adjusting the level of interface complexity. 
\rev{R1-20}{}\rev{R1-17}{Since our user study may not capture the full diversity of all potential users} \rev{R1-19}{and was conducted in a limited amount of time, further investigation, such as a longitudinal study, would help ensure ecological validity in the complexity reduction of visualization and storytelling tools across different users.}

The wide range of domain experts we interviewed suggests that \toolname has potential for diverse applications. However, questions about data preprocessing remain. Expert feedback reveals that real-world network data, such as transactions between companies, communications within and between departments, and distribution channels, are often unstructured and unprocessed. To address this, automated preprocessing of raw data into a compatible format is the next crucial step for \toolname to enhance its versatility.
As a prototype, \toolname does not provide automated methods for extracting interesting patterns from node and link attributes (e.g., anomaly detection for time series~\cite{Shin23}). 
Future work may focus on developing automated methods to improve the utility and generalizability of storytelling tools.
\rev{R1-13}{We also expect the tool can be extended to incorporate datasets from biology (DNA, proteins), software engineering (call graphs, code dependencies, developer activities), and road networks with varying traffic situations.}

\rev{R1-20}{}\rev{R2-1}{
While the dendrogram effectively encodes each snapshot as a leaf node, it faces scalability limitations with large dynamic networks containing numerous snapshots. 
To enhance readability, we need to explore abstraction techniques~\cite{plaisant02, lee06, misue24} such as collapsing and expanding snapshots, despite the additional interaction they introduce.}
\rev{R2-2}{Currently, the prototype supports only a linear cut on the dendrogram. 
Though users can create additional canvases by brushing, enabling multiple cuts at different abstraction levels would offer greater flexibility in generating templates automatically. 
A future system supporting automated story generation with multiple cuts at different levels would improve both scalability and usability for real-world applications.}


\rev{R3-1}{
Besides, examining more sophisticated filtering techniques and community discovery methods would significantly enhance functionality. \toolname offers sliders for intuitive filtering, allowing users to highlight important nodes based on link weights and reduce visual clutter. However, adopting more advanced filtering techniques should be inspected to prevent potential biases~\cite{rocha17} from selecting subgraphs centered on the main character.} 
\rev{R3-2}{The \textit{Community View} displays communities detected by the Louvain algorithm or predefined communities such as affiliation. To enrich narratives, future work could incorporate advanced community discovery methods for temporal networks~\cite{rossetti18}, enabling the tool to suggest more intriguing and comprehensive stories.}


\section{Conclusion} 

This paper presents \toolname, a semi-automatic authoring tool for graph comics based on identifying notable changes in a dynamic graph.
Our approach automatically extracts meaningful events using a causality-preserving hierarchical clustering method. With the algorithmic output, users can adjust granularities by manipulating the aggregation level and refine aesthetics by making editorial decisions down to minute details in comic strips through an intuitive user interface. 
Our user evaluation and expert feedback demonstrate the usefulness of the system, providing promising areas for future work.

\bibliographystyle{abbrv-doi-hyperref}
\section*{Acknowledgments}

This work was supported by the National Research Foundation of Korea (NRF) grant funded by the Korea government (MSIT) (No.RS-2023-00218913, No. 2021R1A2C1004542), by a grant of the Korea Health Technology R\&D Project through the Korea Health Industry Development Institute (KHIDI), funded by the Ministry of Health \& Welfare, Republic of Korea (grant number:HI22C0646), and by the Institute of Information \& Communications Technology Planning \& Evaluation (IITP) grants (No. 2024-00360227-Leading Generative AI Human Resources Development, No. 2020-0-01336–Artificial Intelligence Graduate School Program, UNIST).
Niklas Elmqvist was funded by Villum Investigator grant VL-54492 by Villum Fonden.

\bibliography{dg-comics}

\end{document}